# The SChISM study: Cell-free DNA size profiles as predictors of progression in advanced carcinoma treated with immune-checkpoint inhibitors

Linh Nguyen Phuong[1,2], Frédéric Fina[2,5,6], Laurent Greillier[1,2,4,5], Pascale Tomasini[4,5], Jean-Laurent Deville[5], Romain Zakrajsek[1,2], Lucie Della-Negra[1,2], Audrey Boutonnet[6], Frédéric Ginot[6], Jean-Charles Garcia[6], Sebastien Benzekry[1,2,*], Sebastien Salas[1,2,5,*]

[1] COMPutational pharmacology and clinical Oncology, Centre Inria d'Université Côte d'Azur

[2] Cancer Research Centre of Marseille, Insitut Paoli-Calmettes, Inserm UMR1068, CNRS UMR7258, Aix Marseille University UM105, Marseille, France

[3] ID-Solutions oncology, Marseille, France

[4] Multidisciplinary Oncology & Therapeutic Innovations Department

[5] Assistance Publique—Hôpitaux de Marseille, Timone Hospital, Aix Marseille University, Marseille, France

[6] Adelis Technologies, Labège, France

* = joint senior authors

## Statement of translational relevance

Immunotherapy has transformed cancer care, yet most patients with advanced carcinomas still progress under immune checkpoint inhibitors (ICIs). Robust, noninvasive biomarkers are critically needed to guide treatment decisions beyond PD-L1 immunohistochemistry. In this prospective real-world study, we demonstrate that circulating cell-free DNA (cfDNA) size profiles, quantified via the standardized BIABooster™ device, provide powerful predictive information across multiple tumor types. Long cfDNA fragments (>1650 bp) independently associates with reduced risk of early progression and longer progression-free survival, outperforming PD-L1 and neutrophil-to-lymphocyte ratio (NLR), well established prognostic markers. These associations remained significant after adjustment for confounders and validated across homogeneous subpopulations. Bootstrap analysis confirmed robustness and higher positive predictive value than PD-L1 and NLR. Beyond prediction, our work suggests a biological link between long cfDNA fragments, antitumor immunity, and potential involvement in cGAS-STING pathway activation. This scalable fragmentome-based approach offers immediate translational value for treatment stratification and advancing personalized immuno-oncology.

# ABSTRACT

*Background*

Many advanced cancer patients experience progression under immune-checkpoint inhibitors (ICIs). Circulating cell-free DNA (cfDNA) size profiles offer a promising noninvasive multi-cancer approach to monitor and predict immunotherapy response.

*Methods*

In the SChISM (Size CfDNA Immunotherapy Signature Monitoring) study (NCT05083494), pre-treatment plasmatic cfDNA size profiles from 126 ICI-treated advanced carcinomas were quantified using the BIABooster device. Fragmentome-derived variables (concentration, peaks' position, and fragment size ranges) at baseline were analyzed for associations with early progression (EP, progression at first imaging) and progression-free survival (PFS), using logistic and Cox regression models. Bootstrap analysis validated robustness. Additional analyses were performed in homogeneous subpopulations: first-line lung cancer patients (n = 60) and head-and-neck patients treated with Nivolumab (n = 25).

*Results*

Higher cfDNA concentration and high quantities of short fragments (111-240 base pairs (bp)) were associated with poor response, unlike long fragments (> 300 bp). The proportion of fragments longer than 1650 bp demonstrated highest discriminatory power (AUC = 0.73, C-index = 0.69). It was significantly associated with non-EP (odds ratio = 0.39 [95% CI: 0.25-0.62]) and longer PFS (hazard ratio: 0.54 [95% CI: 0.42-0.68]). These associations remained significant when adjusted for confounders (age, sex, Eastern Cooperative Oncology Group performance status, tumor type, and neutrophil-to-lymphocyte ratio) and across both subpopulations. Bootstrap analysis confirmed robustness with mean accuracy of $70.1 \pm 4.17\%$ and positive predictive value of $55.6 \pm 7.37\%$, in test sets.

*Conclusion*

cfDNA size profiles significantly predicted ICI response and anticipate relapse, outperforming the routinely used marker programmed death-ligand 1 immunohistochemistry and reflecting enhanced immune system activation.

Trial registration: (NCT05083494), date of registration: 2021-10-19.

**Keywords:** cell-free DNA; size profile; fragmentomics; liquid biopsy; cancer; biomarker; prediction of treatment response.

# INTRODUCTION

Immunotherapy has become a powerful clinical strategy for treating cancer. The number of immunotherapy drug approvals has been increasing, with numerous treatments in clinical development[1]. While 20-40%[2] of advanced cancer patients achieve long-term response to immune checkpoint inhibitors (ICI), many do not respond, develop primary or secondary resistance, and experience progression within the first three months of treatment. Currently, response to immunotherapy is assessed solely through imaging based on iRECIST (immune Response Evaluation Criteria in Solid Tumors)[3], a static approach performed after treatment initiation and at fixed intervals (e.g., every 6–8 weeks). This fails to capture tumor kinetics and may prolong ineffective therapy. Relying on fixed thresholds, they do not reflect the nonlinear and delayed nature of immune responses, limiting the ability to distinguish true progression from pseudoprogression or to anticipate relapse. Moreover, while biomarkers predicting response, such as Programmed Death-Ligand 1 (PD-L1) protein expression and Tumor Mutation Burden (TMB), have been widely studied, they are still subject to ongoing standardization efforts and vary across tumor types and patient populations[4,5]). This highlights the urgent need for robust, early predictive biomarkers that are non-invasive, easy to implement, and able to dynamically capture tumor heterogeneity to guide clinical decisions.

In this context, circulating cell-free DNA (cfDNA) sampling as a "liquid biopsy" has been demonstrated to provide clinically relevant information. CfDNA consists of DNA fragments released by various cell types into body fluids, such as blood[6]. It primarily originates from blood[7]—white blood cells, erythrocyte progenitors, vascular endothelial cells[8]—and tumor cells, being released under necrosis, apoptosis[9], lysis[10], or active secretion of exosomes[11]. With a short half-life (15 minutes-2 hours[12]), cfDNA reflects real-time system activity. In cancer patients, the plasmatic concentration of global cfDNA — shed by both wild-type and tumor cells — rises and fluctuates with tumor size, number of metastases, and treatment[13,14].

Ninety percent of cfDNA studies in immuno-oncology have focused on circulating tumor-derived cell-free DNA (ctDNA) and its sequencing. They established predictors of response to immunotherapy by individually detecting the presence of somatic genomic alterations (mutations, TMB, and microsatellite instability) as biomarkers for monitoring treatment. However, even for cancers with a high mutation rate, approximately twenty percent of patients do not have ctDNA traceable anomalies[15].

In addition, wild-type cfDNA is primarily shed by immune cells[8,9]. Therefore, interest has emerged for features of global cfDNA, beyond ctDNA. In 2015, Ivanov et al. showed that fragment patterns differed across cancer patients and tissue of origin[16]. Several studies have since investigated various fragmentation features, such as end, breakpoint or jagged ends motifs, nucleosome footprints, and fragment sizes. Since cancer patients appear to have an enrichment in shorter fragments (90-150 bp), focusing on the cfDNA fragment size could enhance the biology of cfDNA release[17]. Then the analysis of cfDNA fragments size distribution[18] offers a non-invasive method for assessing treatment response independently of a specific molecular target, cancer type, and treatment[18].

The ongoing proof-of-concept study SChISM—*Size cfDNA Immunotherapy Signature Monitoring*—introduces a novel approach leveraging a patented and standardized method for cfDNA quantification (BIABooster[19,20], Adelis technologies, France). SChISM aims to investigate the predictive potential of cfDNA fragment size profiles for identifying resistance to immune checkpoint inhibitors (ICI). Specifically, we evaluated the predictive and prognostic performance of pre-treatment cfDNA size profiles in routine care patients to anticipate early progression (EP) and progression-free survival (PFS) following immunotherapy across four carcinoma cancer types, non-small cell lung cancer (NSCLC), head and neck squamous cell carcinoma (HNSCC), urothelial carcinoma (UC), and clear cell renal cell carcinoma (ccRCC). The goal is to enable early therapeutic adaptation, prevent ICI-related disease progression, with cfDNA fragment sizes as potential universal biomarkers.

## METHODS

### *Study design and patient population*

SChISM (Size CfDNA Immunotherapy Signature Monitoring) is a prospective, multicenter, collaborative, non-interventional clinical study with minimal risk, conducted within several departments of the Assistance Publique–Hôpitaux de Marseille (AP-HM). CfDNA analysis was performed using BiaBooster™ technologies to investigate predictive and prognostic value in 337 patients with metastatic and/or recurrent NSCLC, HNSCC, ccRCC, and UC, initiating treatment with ICIs as standard care since 2021. Recruitment spanned two years across four oncology departments. Follow-up extended until progression under ICI therapy or 12 months of treatment. cfDNA samples were collected immediately before each ICI-infusion until the first imaging evaluation, then every two cycles for one year, and at recurrence. Sampling followed a standardized schedule across tumor types. The final analytic cohort included baseline data from patients meeting inclusion criteria with adequate follow-up. Exclusions were melanoma diagnosis (n = 110), non-evaluable disease status (n = 43), and insufficient follow-up data (n = 58).

### *Clinical and biological data*

Clinical and biological data were prospectively collected. Variables included age, sex, Eastern Cooperative Oncology Group (ECOG), tumor type, and disease stage immunotherapy initiation. Treatment response was assessed every three months with contrast-enhanced computed tomography (CT) scans, interpreted using RECIST 1.1 or iRECIST criteria. Sum of largest diameters (SLD) was computed from these CT scans. Dates of radiological progression, last follow-up, or death were recorded for survival analyses. Routine laboratory parameters collected at baseline and during follow-up, included complete blood count, serum lactate dehydrogenase (LDH), serum creatinine, and liver function tests (AST, ALT, ALP, and bilirubin). LDH and neutrophil-to-lymphocyte ratio (NLR) were selected as nonspecific biomarkers that have been extensively studied as both a prognostic and predictive factor in the context of immunotherapy[21,22]. PD-L1 expression scores, including Combined Positive Score (CPS) and Tumor Proportion Score (TPS), were collected when available in routine practice. Data were entered into a secured, anonymized database and regularly monitored.

*Endpoints*

Clinical endpoints included EP, defined as confirmed progression at the first radiological assessment (3 months after treatment initiation). PFS was defined as time from first infusion to the date of progression, death or last follow-up, whichever occurred first. Progression date was defined as the date of disease progression, or the date of last follow-up.

*Regulatory and Ethical Aspects*

This non-interventional study followed the Declaration of Helsinki, Good Clinical Practice, and French regulations. Given the non-interventional nature and minimal risks associated with cfDNA sampling during routine visits, approval was sought from the ethics committee (Comité de Protection des Personnes) and registration completed with the French Ministry of Health under the Jardé Law (Category 2 research).

The sponsor (AP-HM) oversaw data protection regulations (GDPR) compliance through secure, anonymized electronic Case Report Forms. Written informed consent was obtained before any procedure.

*Blood collection and cfDNA quantification*

The certified AP-HM Biobank (CRB, NFS 96-900, ISO 9001:2015) managed plasma processing and storage. Peripheral blood (Roche cfDNA Collection Tube) was centrifuged at 1,600×g for 10 minutes at room temperature. Plasma supernatant was collected into a 15 mL conical tube and subjected to a second centrifugation at 4,500×g for 10 minutes to remove residual cellular debris. Resulting plasma was transferred into a new 15 mL tube and aliquoted in 300 to 500 µL volumes, then stored at –80°C. Plasma was analyzed at Adelis (Labège, France), using BIABooster for size profile characterization of cfDNA.

**Reagents:** The cfDNA kit (Adelis, # 16-BB-cfDNA, France) was used.

**Instrument and capillary assembly:** Experiments were carried out with a G7100A CE system (Agilent Technologies, Germany) equipped with a Zetalif fluorescence detector (Adelis, France) and a BIABooster capillary device (Adelis, # 16-BB-DNA/11, France). DNA was concentrated at the junction of two capillaries with different diameters using dual hydrodynamic and electrokinetic actuation. During this concentration step, the salts and proteins contained in the cfDNA sample were removed, allowing for in-line purification, enabling size-dependent migration. Gel electrophoresis quantified fragments, providing cfDNA size distribution (100-1500 bp) with 10 fg/μL sensitivity[20].

**BIABooster DNA analysis:** Plasma was pretreated with lysis buffer (56°C, 2h, 900rpm) to release cfDNA from vesicles and protein complexes. One μL was injected into the BIABooster device for cfDNA analysis. The BIABooster device records fluorescence intensity as DNA fragments migrate through the fluorescence detector, creating curves that show fluorescence vs. time (in minutes, Figure 1A). The BIABooster Analytics software converts fluorescence into concentration and time into fragment size, based on a DNA ladder. The device gives reliable size and concentration measurements for DNA fragments between 75 and 1650 base pairs (bp). Below 75 and above 1650 bp, all fragments migrate at similar speeds, making size determination unreliable. However, the values do still reflect the relative amounts of fragments at these ranges. They are still useful for sample comparisons, but don't represent true sizes or concentrations.

*cfDNA variables*

One cfDNA curve is defined as the concentration of fragments in pg/μL according to the fragment size in bp. Twelve quantitative variables were derived from each curve, based on range of sizes corresponding to the length of DNA strands wrapped around nucleosomal multiples (Figure 1B):

- The first peak's position (in bp) denoted as $P_1$, corresponding to the most frequent size of the fragments originating from mono-nucleosomes.

- The second peak's position (in bp) denoted as $P_2$, corresponding to the most frequent size of the fragments originating from di-nucleosomes.

- The difference between $P_2$ and $P_1$, denoted as $P_2 - P_1$.

- The left half-width of the first peak at mid-height (in bp), denoted as $HW$.

- The global concentration $C_{TOT}$, defined as the area under the cfDNA curve between 75 and 1650 bp computed by the trapeze method.

- The relative concentration in pg/μL of cfDNA fragments measuring between $x_i$ and $x_j$ bp, to the total concentration $C_{TOT}$, defined as the following absolute concentration

$$R_{[x_i,x_j]} = \frac{C_{[x_i,x_j]}}{C_{TOT}}$$

  for $[x_i, x_j] \in \{[75,111], [111,240], [240,370], [370,580], [580,1650]\}$.

- The relative quantity of cfDNA fragments of less than 75 bp to the total concentration $C_{TOT}$, in relative arbitrary unit (r.a.u.), denoted as $R_{<75}$.

- The relative quantity of cfDNA fragments over 1650 bp to the total concentration $C_{TOT}$, in r.a.u, denoted as $R_{>1650}$.

*Preprocessing*

We assessed the normality of variable. For a given feature $X_j$, we fitted both normal and log-normal distributions and computed the associated Bayesian Information Criterion (BIC), $BIC_{normal}^j$ and $BIC_{log-normal}^j$. For features $X_j$ containing negative or zero values, we applied a shift of $\epsilon - \min_j x_j$, with $\epsilon = 0.01$. Features satisfying $BIC_{normal}^j - BIC_{log-normal}^j \geq 6$ were log-transformed.

Log-transformed features included NLR, continuous TPS, $R_{>1650}$, $C_{TOT}$, $R_{[75,111]}$, $R_{[580,1650]}$, $R_{>1650}$, $HW$, and $R_{<75}$.

All the continuous features were scaled within the logistic and Cox regressions. Patients with missing values were removed.

*Statistical analysis*

All statistical computations used R version *4.4.0*. Pearson's Chi-squared and Welch t-tests compared respectively categorical and continuous features with EP status. Associations with EP and PFS were assessed with logistic (*stats::glm 4.4.0*) and Cox regression (*survival::coxph 3.8.3*), first univariable (UV) before then adjusted (MV) for clinical confounding features: age, ECOG status, tumor type, and NLR. Survival was displayed with Kaplan-Meier curves (*survival::survfit 3.8.3*). Continuous features were optimally dichotomized using *survminer::surv_cutpoint 0.5.0*, maximizing the log-rank statistic while ensuring a minimum of 20% of patients per group.

Analysis were repeated in two homogeneous subpopulations defined by treatment line and cancer type: first line NSCLC patients[23] (n = 60), and HNSCC patients treated with Nivolumab in monotherapy, corresponding to HNSCC patients resistant to platinum-based chemotherapy[24] (n = 25). Both subpopulations represented the two largest homogeneous subgroups in the overall study population. In the first subpopulation, continuous TPS PD-L1 was added as confounder in the MV analysis, being available for 54 patients. In the analysis of the second subpopulation, CPS PD-L1 could not be added as confounder, being available for only 10 patients.

For cfDNA predictive performance, we investigated optimal threshold by performing 70% bootstrap resampling with replacement. Thresholds were optimized via the distance between receiver operating characteristic (ROC) curve and the point (0,1) (*cutpointr* 1.2.0). Accuracy, specificity, sensitivity, positive predictive value (PPV), and negative predictive value (NPV) were computed on both training and out-of-bag datasets, repeated 100 bootstraps. This robustness framework was not applied to cohorts A and B due to limited size.

Unsupervised hierarchical clustering was performed on the multi-cancer cohort using Euclidian distance and Ward linkage method to cluster both patient and features (cfDNA features, NLR, LDH, TPS, and age), using the package *ComplexHeatmap* 2.22.0.

# RESULTS

## *Patient characteristics*

The data comprised 126 patients with advanced/metastatic carcinoma: NSCLC (n = 68), UC (n = 10), HNSCC (n = 34), and ccRCC (n = 14). Mean age was 64 years, with 73% of male (Table 1). Patients received first-line (n = 100), second-line (n = 25), or third-line immunotherapy (n = 1): nivolumab (n = 38), pembrolizumab (n=87), or atezolizumab (n = 1). Forty-eight patients has concomitant chemotherapy (43 NSCLC, 5 HNSCC, all first-line), 11 had concomitant targeted therapy (9 ccRCC, 1 UC, 1 HNSCC), and 2 ccRCC received nivolumab in combination with ipilimumab.

PD-L1 CPS was available for 56% of HNSCC patients and categorized into three levels: < 1, 1-19, and ≥ 20%[25]. TPS was available for 91% of the NSCLC patients and categorized into three levels: < 1%, 1-49, and ≥ 50%[26]. Pre-treatment LDH was available for 56%, SLD measurements 92%, and NLR values for 87% of the cohort.

Median PFS was 9.53 months [95% confidence interval (CI): 6.8-NA] (Table 1), with follow-up of 16 months [95% CI: 13.6-21.5]. Disease progression occurred in 69 patients (55%), with 44 patients (34.9%) progressing within three months of treatment. PFS varied by tumor type (Figure S1). EP proportion was 59% in HNSCC, 70% in UC, and 21% in NSCLC/ccRCC patients (Figure S1). Median overall survival (OS) was 14.4 months [95% CI: 12-NA] (Figure S2).

Pre-treatment cfDNA size distribution revealed a median primary peak at 157 bp (range: 132-160) (Table S1, Figure S3), and secondary peak at 307 bp (range: 274-326), reflecting mono- and dinucleosomes (Figure 1B-C). A median 58% (range: 28-85) of fragments were 111-240 bp. Considerable inter-patient variability was observed across all fragment sizes (Figure 1C). cfDNA concentration was similar between EP and non-EP across fragments lower than 580 bp, while fragments between 580 and 1650 bp represented 8% (range: 1-17) of the total cfDNA for EP, and 11% (range: 2-36) for non-EP patients (Figure S3). This variability was hardly visible in the concentration vs. size curves (Figure 1B) because of the concentration scale but was apparent in the fluorescence vs. time curves (Figure 1A). Similarly, high-molecular weight (HMW, > 1650 bp) fragments averaged 0.031 r.a.u. (range: 0.005-0.086) in EP patients, versus 0.05 r.a.u. (range: 0.006-0.165) in non-EP patients.

## *cfDNA fragmentome is associated with response across cancer types*

Across all patients, higher cfDNA concentration $C_{TOT}$ was positively associated with EP (AUC = 0.65, Table 2A) and PFS (C-Index = 0.62). Patients with $C_{TOT} > 14$ pg/μL had significantly shorter PFS (Figure 2A). In MV analysis, $C_{TOT}$ remained significantly associated with EP (OR = 2 [95% CI: 1.1-3.4], $p = 0.006$) and PFS (HR = 1.8 [95% CI: 1.3-2.3], $p < 0.0001$).

Short fragments $R_{[111,240]}$ was associated with EP in UV analysis (OR = 1.5 [95% CI: 1-2.2], $p = 0.03$, AUC = 0.6, Table 2A, Figure 2A), but was no longer associated after adjusting for confounding factors. However, $R_{[111,240]}$ was strongly associated with shorter PFS in both UV and MV analysis (MV HR = 1.7 [95% CI: 1.3-2.4], $p = 0.0004$). Conversely, cfDNA variables related to fragments longer than 370 bp were associated with better response to ICIs. $R_{>1650}$ showed strongest performance, achieving an AUC of 0.73 and a C-index of 0.69. Higher quantity of the longest fragments was significantly associated with non-EP (t-test: $p < 0.0001$) and longer PFS (log-rank test: $p < 0.0001$) (Table 2A, Figure 2A), both in UV and MV analyses (MV OR = 0.28 [95% CI: 0.13-0.6], $p = 0.0009$; MV HR = 0.44 [95% CI: 0.33-0.58], $p < 0.0001$). Similar observations were found for $R_{[580,1650]}$ (AUC = 0.65, C-Index = 0.64). An increase of 1% in the relative amount of $R_{[580,1650]}$ was associated with a 50% reduction in both odds and hazards of progression, both in the UV and MV analyses. The difference between first and second peak's location showed the second best performance (AUC = 0.72, C-index = 0.67). It was associated with non-EP (t-test: $p < 0.0001$) and longer PFS (log-rank test: $p < 0.0001$) (Table 2A, Figure 2A), both in UV and MV analyses (MV OR = 0.36 [95% CI: 0.18-0.73], $p = 0.005$; MV HR = 0.6 [95% CI: 0.44-0.83], $p = 0.002$). Additionally, $P_2$ was associated with non-EP (AUC = 0.64) in UV and MV (OR = 0.46 [95% CI: 0.25-0.85], $p = 0.01$) analyses, and a longer PFS in UV and MV (HR = 0.59 [95% CI: 0.44-0.8], $p = 0.0006$) analyses (Table 2A, Figure 2A). Several ratios of cfDNA features were investigated. The ratio $\frac{R_{>1650}}{R_{[111,240]}}$ achieved the best performance for association with EP, with an AUC of 0.71, whereas the ratio $\frac{R_{[580,1650]}}{R_{[75,111]}}$ achieved the best performance for association with the PFS, with a C-index of 0.66.

Among clinical and biological variables, ECOG status, sex, and NLR were significantly associated with EP and PFS (Table 2A). NLR demonstrated an AUC of 0.69 and was strongly associated with EP (UV OR: 2.4 [95% CI: 1.5-4], $p = 0.0006$) and shorter PFS (UV OR: 1.6 [95% CI: 1.2-2.2], $p = 0.002$, Figure S4). Patients with an ECOG status ≥ 2 had a markedly higher risk of EP (UV OR: 6.0, 95% CI: 2.1–17, p = 0.0009) and significantly shorter PFS (UV HR: 3.3, 95% CI: 1.8–6.2, p = 0.0001).

Weak correlations were observed between cfDNA variables and five classical predictors of progression (age, ECOG, NLR, LDH and SLD, Figure 2B), suggesting orthogonal added predictive value of cfDNA fragmentome. Notably, correlations with NLR were all very low (max $|r| = 0.14$). Among patients with available NLR data (87%), $R_{>1650}$ and $P_2 - P_1$ additionally identified 11 out of 37 (22%) EP patients and 18 out of 73 (21%) non-EP patients missed by NLR (Figure 2C).

The predictive ability of cfDNA variables was assessed using a bootstrap analysis (Figure 2D). $R_{>1650}$ achieved the best performance across both datasets. In training sets, accuracy was $73 \pm 4.57\%$, with specificity of $72.1 \pm 6.01\%$ and sensitivity of $74.8 \pm 7.21\%$. PPV was $59.7 \pm 7.76\%$ and NPV was $83.7 \pm 5.46\%$. In out-of-bag datasets, accuracy was $70.1 \pm 4.17\%$, specificity $70.3 \pm 6.50\%$, sensitivity $69.5 \pm 11.3\%$, PPV $55.6 \pm 7.37\%$, and NPV $81.5 \pm 5.19\%$.

$P_2 - P_1$ showed less robustness, with PPV of $55.0 \pm 8.42\%$ and $46.9 \pm 7.47\%$ in the train and test datasets, respectively.

### *Unsupervised learning reveals cross-cancer patient clusters associated with progression*

Unsupervised hierarchical clustering (Figure 3) identified three main feature clusters, defined by their cfDNA size: 1) total cfDNA concentration ($C_{TOT}$), and mononucleosome-associated fragment features ($R_{[111,240]}$), grouped with LDH, NLR, age, and TPS, 2) very short fragments ($HW$, $R_{<75}$ and $R_{[75,111]}$), and 3) long fragments ($R_{[240,370]}$, $R_{[370,580]}$, $R_{[580,1650]}$, and $R_{>1650}$) and peak's variables ($P_1$, $P_2$, and $P_2-P_1$). Notably, conventional variables, particularly NLR, exhibited distinct variation patterns from cfDNA features.

Three patient clusters (termed A, B, C) emerged with different fragmentome profiles. Although the clustering was agnostic to the EP endpoint, these clusters were significantly associated with EP ($p = 0.04$, Pearson's Chi-squared test). The largest one (cluster A, n = 24) was characterized by elevated cfDNA concentrations and a predominance of short fragments over long ones. In this cluster, 54% of the patients experienced EP. Cluster B (n = 17) was characterized by elevated quantities of very short fragments (<111 bp). It was associated with good prognosis, with 82% of patients not experiencing EP (compared to 65% overall). Cluster C (n = 85) was characterized by elevated quantities of long fragments and low total cfDNA levels, and was associated with non-EP (67%). All tumor types were represented aross the three clusters.

### *In NSCLC first-line subgroup, long cfDNA fragments were more strongly associated with non-EP and longer PFS*

In NSCLC first-line subgroup (n = 60), associations were similar but with stronger effect. Five cfDNA variables achieved AUC ≥ 0.83 and C-Index ≥ 0.71 ($P_2$, $P_2 - P_1$ $R_{>1650}$, $C_{TOT}$, and $R_{[370,580]}$, Table 2B). $P_2$ demonstrated the strongest predictive performance, with an AUC of 0.92 and significant negative association with EP in UV analysis (OR = 0.067, [95% CI: 0.011-0.41], $p = 0.003$). $P_2 - P_1$ also showed high performance (AUC = 0.82) and was significantly associated with non-EP (UV OR = 0.19 [95% CI: 0.059-0.61], $p = 0.005$). $R_{>1650}$ achieved an AUC of 0.84 and was significantly associated with non-EP (UV OR = 0.22 [95% CI: 0.09-0.56], $p = 0.001$). Higher cfDNA concentration was associated with higher risk of EP (AUC = 0.830, UV OR = 4 [95% CI: 1.6 - 9.9], $p < 0.01$). However, none of these associations remained significant in MV analysis.

$P_2$, $P_2 - P_1$, $R_{>1650}$, and $C_{TOT}$ were significantly associated with PFS with and without adjusting for confounders. $P_2$ was associated with longer PFS (MV HR = 0.4 [95% CI: 0.23-0.7], $p = 0.001$, C-Index = 0.76, Figure 2A). $P_2 - P_1$ was associated with longer PFS (MV HR = 0.47 [95% CI: 0.26-0.85], $p = 0.01$, C-Index = 0.75). $R_{>1650}$ was also associated with longer PFS (MV HR = 0.37 [95% CI: 0.21-0.67], $p = 0.0009$). In contrast, $C_{TOT}$ was associated with shorter PFS (MV HR = 1.9 [95% CI: 1.2-3.1], $p = 0.009$). Among the clinical and biological variables, NLR achieved the best performance ( AUC = 0.79) and was outperformed by $P_2$, $P_2 - P_1$, $R_{>1650}$, and $C_{TOT}$.

While correlations between cfDNA and NLR remained weak ($|r| < 0.35$), stronger correlations were observed with LDH in this cohort (Figure 4B), possibly due to the small number of patients with LDH available (n = 16). No strong correlation was observed with TPS PD-L1 (n = 55, $|r| < 0.32$). $R_{>1650}$ and $P_2$ identified 1 out of 7 (14%) EP patients and 13 out of 45 (29%) non-EP patients missed by NLR. In contrast, NLR only identified 1 additional non-EP patient missed by $R_{>1650}$ or $P_2$ (Figure 4C).

***In HNSCC nivolumab subgroup, HMW fragments remained significantly associated with longer PFS***

In HNSCC nivolumab subgroup, most of cfDNA variables showed no association with EP, possibly due to the smaller sample size (n = 25). Nevertheless, high $R_{>1650}$ remained associated with longer PFS in UV analysis and was still significant when adjusting with confounders (HR: 0.41 [0.23-0.75], $p = 0.003$, Table 2C, Figure 4D). Patients with a baseline $R_{>1650}$ lower than 0.0036 r.a.u. achieved a median PFS of 2.07 [95% CI: 1.8-3.1], against 5.57 [95% CI: 1.9-NA] for the remaining patients (log-rank test $p = 0.0003$). NLR outperformed $R_{>1650}$ (AUC = 0.8).

In this cohort, $R_{>1650}$ was moderately associated with SLD (n = 22, $r = -0.34, p = 0.12$) and CPS PD-L1 (n = 10, $r = -0.43, p = 0.22$) (Figure 4E). Three out of 16 EP patients and 1 out of 7 non-EP were correctly identified by $R_{>1650}$ and not NLR, while conversely only one EP and 2 non-EP patients were specifically associated by NLR and not $R_{>1650}$ (Figure 4F).

Optimal threshold for $R_{>1650}$–computed to achieve the best separation of PFS curves– remained consistent across all three cohorts, ranging between 0.036 and 0.039 (overall feature range: 0.005 to 0.16). The optimal $P_2$ was consistent between multi-cancer and lung cohorts. Conversely, $C_{TOT}$ threshold was approximately 14 pg/µL in both multi-cancer and HNSCC cohorts, but doubled in the lung cohort, due to extreme outlier values (two lung patients showing $C_{TOT} > 400$ pg/µL).

## DISCUSSION

This study provides a proof-of-concept, prospective, real-world evidence supporting the utility of cfDNA fragmentomics—specifically cfDNA size profiles—as an innovative, non-invasive, pre-treatment biomarker to early predict response to ICIs across four advanced solid malignant carcinomas. The most predictive cfDNA variable, quantifying fragments greater than 1650 bp, was associated with both lower odds of EP and prolonged PFS, and high-performed conventional clinical and biological features. These associations remained significant after adjusting for established confounders and were validated through bootstrap-based internal validation, supporting their statistical robustness and reproducibility. Importantly, the predictive and prognostic value of this cfDNA signal remained significant in first-line NSCLC patients and were significantly associated with PFS in platinum-resistant HNSCC patients, highlighting its consistency within homogeneous clinical subgroups, regardless of histological subtype or treatment context. These findings suggest that cfDNA size profile appears to capture biologically relevant information not reflected in current clinical or pathological markers. Notably, cfDNA variables showed minimal correlation with common variables such as LDH, PD-L1 expression (TPS/CPS PD-L1), NLR, and tumor burden (SLD), reinforcing their role as independent indicators of immune response. Particularly, it identified a subset of patients at risk for EP who were not detected by NLR, highlighting the potential of integrating cfDNA fragmentomics data into multivariable models and clinical decision-making algorithms, to improve early treatment stratification.

However, a key challenge remains the accurate prediction of disease progression under treatment. Current standard biomarkers show limited predictive performance in multi-cancer contexts, with TMB and PD-L1 achieving pooled AUCs of 0.69 and 0.65 and pooled PPVs of 0.42 and 0.34[27], respectively. In contrast, our cfDNA-based classifier demonstrated superior performance for prediction. Within the SChISM study specifically, cfDNA fragment size profiling outperformed other established biomarkers—including NLR[28,29] and PD-L1 status—for predicting treatment response.

Previous research has mainly focused on either the total cfDNA concentration or the abundance of short fragments (< 200 bp)[30]. While elevated total cfDNA levels have been associated with shorter PFS and OS[31,32], size-based cfDNA variables have been limited to diagnostic applications[17] or chemotherapy response prediction, where shorter fragments showed predictive value[30]. Healthy individuals typically exhibit cfDNA size distributions ranging from 130 to 200 base pairs (bp), with a peak at 166 bp[33]. This phenomenon seems to be linked to the nucleosome footprint (nucleosome positions)[34]. This specific length corresponds to DNA fragments wrapped around a nucleosome core (~147 bp) plus connecting linker sequence of ~20 bp[35], representing the preferred cleavage sites where nucleases cut DNA during apoptosis[16]. These fragments appear to originate from a double process of apoptosis-induced proliferation and proliferation-induced apoptosis[9]. In contrast, longer fragments such as 1,000 bp fragments appear to originate from necrosis or phagocytosis of necrotic tumor cells by macrophages[36], and are much less present in the plasma[9]. Additionally, cancer patients appear to have an enrichment in shorter fragments (90-150 bp)[17]. In contrast, the SChISM study revealed a distinct pattern for immunotherapy response prediction. The analysis demonstrated that longer fragments—specifically di-nucleosomal ($P_2$) and long fragments ($R_{[580,1650]}$ and $R_{>1650}$)—more effectively discriminate between progressors and responders compared to mono-nucleosomal fragments ($R_{[111,240]}$, $P_1$) or total cfDNA levels ($C_{TOT}$). Furthermore, the interval between the first two peak locations exhibited high predictive performance. These finding suggest that the biological processes driving DNA fragmentation could reflect the pre-treatment immune system status of patients. Different ratios of long to short fragments were explored, in line with the concept of cfDNA integrity[37] (cfDI), which computes ratios on specific genetic locus, and is typically assessed for diagnostic or prognostic purposes. However, relying on the single cfDNA feature $R_{>1650}$ consistently yielded better performance than either ratio.

Mechanistically, the enrichment of long cfDNA fragments in responders could result from activation of immune cells[38], but also may reflect tumor cell death[39]. It may suggest increased DNA release through processes such as necrosis, NETosis, phagocytosis, or active secretion[40]. Patients at higher risk of progression exhibit shorter dinucleosome-derived fragments ($P_2$ measuring less than 302 bp) as well as nucleosomes more tightly packed together (interval between $P_2$ and $P_1$ lower then 154 bp). These both observations can be due to closer nucleosome spacing, altered nuclease activity[41], hypomethylation, or altered proliferation[9]. Overall, the observed size profile indicates that higher baseline cfDNA fragmentation may be prognostic of progression within the first three months of treatment and shorter PFS. From a causal perspective, cfDNA may play an active role in modulating immune responses. Long cfDNA fragments have the potential to activate the cGAS-STING pathway[42,43], triggering interferon-mediated immune activation. This hypothesis is supported by emerging preclinical evidence showing that cytosolic double-stranded DNA can act as an immunostimulant and enhance sensitivity to ICIs[44]. In contrast to short apoptotic fragments generated by orderly nucleosomal cleavage, associated with a tumor-derived origin[45], long fragments may function as danger-associated molecular patterns (DAMPs), promoting antitumor immunity[43]. Therefore, lower cfDNA fragmentation at baseline, in combination with ICI therapy, may facilitate more effective tumor eradication. These findings not only introduce a novel predictive biomarker but also offer mechanistic insights into resistance and potential new actionable target for ICI-resistant patients[39].

CfDNA size profile offers practical advantages for clinical implementation. Because it contains fragments from both tumor and wild-type cells, global cfDNA is independent of tumor genotype, making it applicable across genetically heterogeneous tumors. It also bypasses the need for tissue biopsies, enabling non-invasive serial monitoring. Moreover, using the BIABooster™ device, cfDNA concentration and size can be quantified without prior DNA extraction, facilitating fast and standardized implementation in clinical laboratories. These advantages, combined with the prognostic relevance of cfDNA size, support the development of new tools to guide early treatment decisions—particularly to avoid the unnecessary continuation of ICI therapy in sensitive patients after the induction phase, thereby reducing treatment-related toxicity and associated costs. If future studies in more homogeneous cohorts confirm a higher PPV, such a test could also help avoid initiating unnecessary treatment in poor responders. Ultimately, its clinical utility will depend on the cancer type, the clinical context, and the expected efficacy of the therapy.

However, while our findings are statistically robust, they require external validation in independent and disease- or treatment-specific cohorts. Moreover, size ranges were arbitrary set according to technology limits and as nucleosomal-length bounds. More optimized intervals could lead to better performance. Furthermore, the analysis is limited to a single pre-treatment timepoint. To address this, we plan to integrate both baseline and on-treatment cfDNA size profiles, collecting plasma samples before each immunotherapy cycle, into a machine learning framework to provide dynamic biomarkers of response.

Dealing with longitudinal data, mechanistic modeling frameworks could offer deeper insights into the biological processes that drive cfDNA fragmentation and their connections to tumor evolution, immune dynamics, and consequently immunotherapy outcomes. Such approaches could simulate key processes including cfDNA release and clearance, immune system activation, and treatment-induced changes in tumor and immune cell turnover, ultimately linking cfDNA size profiles to patient immune responses. When combined with machine learning—an approach referred to as "mechanistic learning"[46]—, these frameworks can yield hybrid predictive algorithms that combine biologically interpretable features with high-dimensional data. Longitudinal monitoring of cfDNA fragment dynamics—especially in the early treatment phases—may improve early response prediction, differentiate pseudoprogression from true progression, and support adaptive therapeutic decisions. Ultimately, validating and refining these models in large, independent cohorts is crucial for translating cfDNA size profiles from a research tool to a clinically actionable biomarker.

In conclusion, our findings highlight cfDNA fragment size profiling—particularly the quantification of HMW fragments—, as a promising, multi-cancer, minimally invasive biomarker for early response prediction to ICIs. The integration of this variable into longitudinal frameworks may pave the way for precision immuno-oncology strategies and real-time adaptive treatment management.

# FIGURES

*Figure 1: Cancer patients exhibit different cfDNA size profiles under ICI treatment*

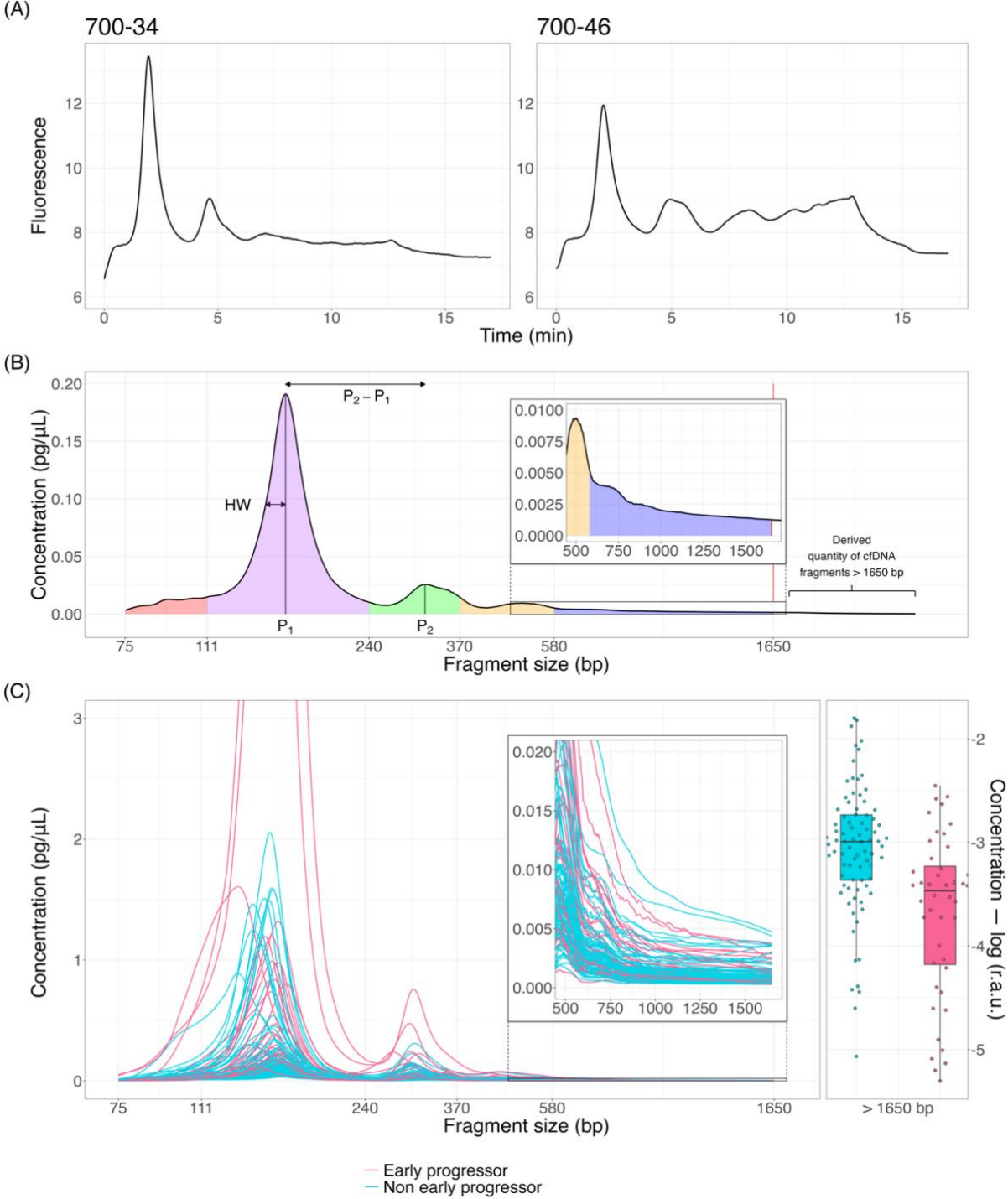

(A) Fluorescence vs. time curve, distinguishing patients with low vs. high levels of high-molecular-weight DNA (700-34 vs. 700-46). (B) Pre-treatment cfDNA size profile of the patient 700-46, derived from the fluorescence/time plot on the right panel of (A). Ten of the extracted cfDNA variables are displayed: - the first, second peak' locations ($P_1$ and $P_2$) in base pairs (bp) and their difference ($P_2 - P_1$); - the half-width of the left side of the first peak $HW$; - the concentration (area under the curve) of cfDNA from different size ranges: $R_{[75,111]}$: red area; $R_{[111,240]}$: purple area; $R_{[240,370]}$: green area; $R_{[370,580]}$: yellow area; $R_{[580,1650]}$: purple area.; the quantity of cfDNA greater than 1650 bp (C) CfDNA data of the 126 patients, colored by outcome: early-progressor (pink), non early-progressor (blue). Left: pre-treatment cfDNA size profiles. Zoom: focus on size distributions between 500 and 1650 bp Right: pre-treatment distribution of $R_{>1650}$.

*Figure 2: Higher ratio of long cfDNA fragments is associated with extended progression-free survival and no early progression under treatment in the multi-cancer cohort*

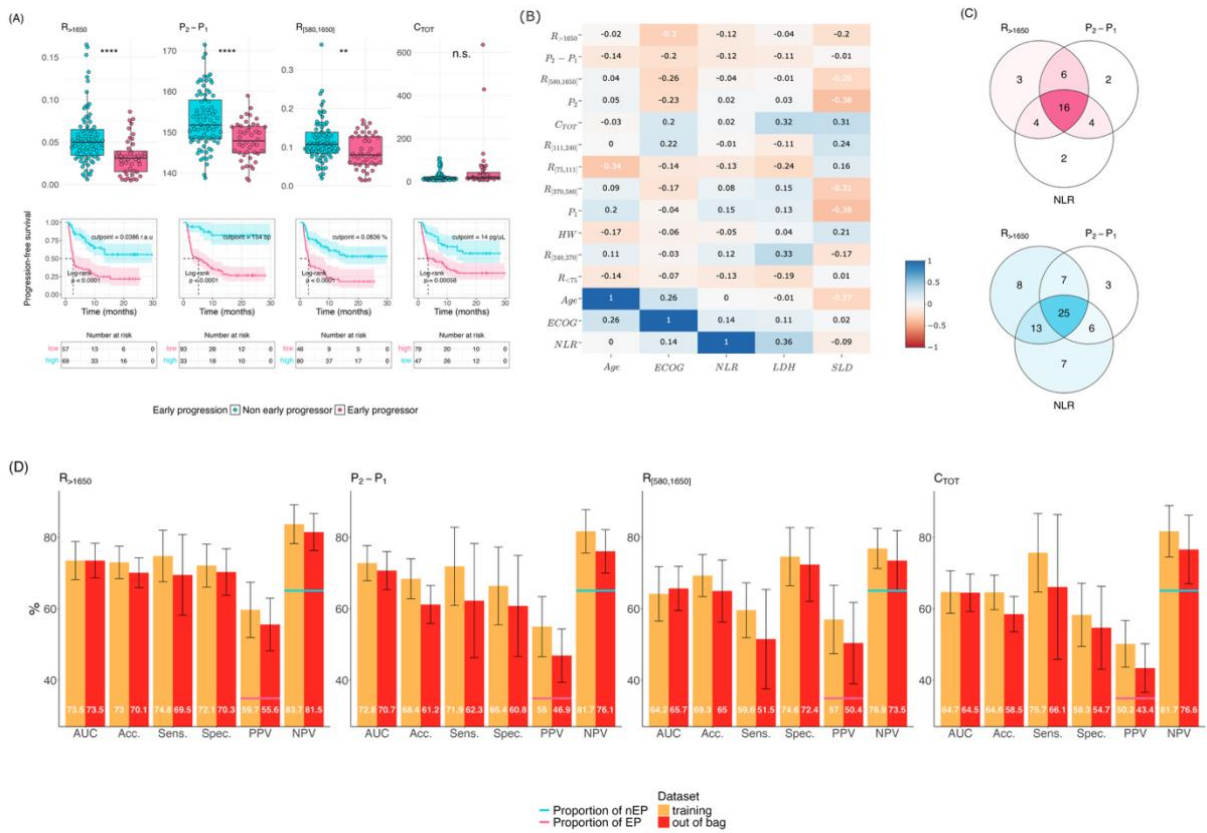

(A) Distribution of the best cfDNA variables ($R_{>1650}$, $P_2 - P_1$, $R_{[580,1650]}$, and $C_{TOT}$) by early progression status and progression-free survival. Boxplots show cfDNA variable distributions stratified by early progression status and display the Welch two sample t-test significance. Kaplan-Meier curves display progression-free survival stratified by cfDNA levels, display the log-rank test p-value and the optimal threshold computed to stratify patients between long and short progression-free survival (see Methods). (B)

Pearson correlation was computed between the eleven cfDNA variables and five conventional features: age, Eastern Cooperative Oncology Group (ECOG) status, neutrophil-to-lymphocyte ratio (NLR), lactate dehydrogenase (LDH), and sum of the largest target diameters (SLD) ($n_{SLD} = 116$, $n_{NLR} = 110$, $n_{LDH} = 71$). (C) Venn diagrams were generated based on the counts of true early progressors (pink) and true responders (blue) identified as the two best cfDNA variables ($R_{>1650}$, $P_2 - P_1$) in the statistical analyses and NLR. (D) Area under the receiver operating curve (AUC), accuracy (Acc.), sensitivity (Sens.), specificity (Spec.), positive predive value (PPV) and negative predictive value (NPV) are displayed for $R_{>1650}$, $P_2 - P_1$, $R_{[580,1650]}$, and $C_{TOT}$. Each performance metric was computed 100 times on both bootstrapped datasets (yellow) and out-of-bag datasets (red). The reported values represent the mean of the 100 iterations, while error bars indicating the mean ± 1 standard deviation. The true rate of EP is indicated in pink within the PPV bars, and the true rate of non-EP is indicated in blue within the NPV bars.

(Significance: ****: p-value < 0.0001; ***: p-value < 0.001; **: p-value < 0.01; *: p-value < 0.05; n.s.: non-significant, > 0.05)

*Figure 3: Patients are clustered according to their fragment size distribution*

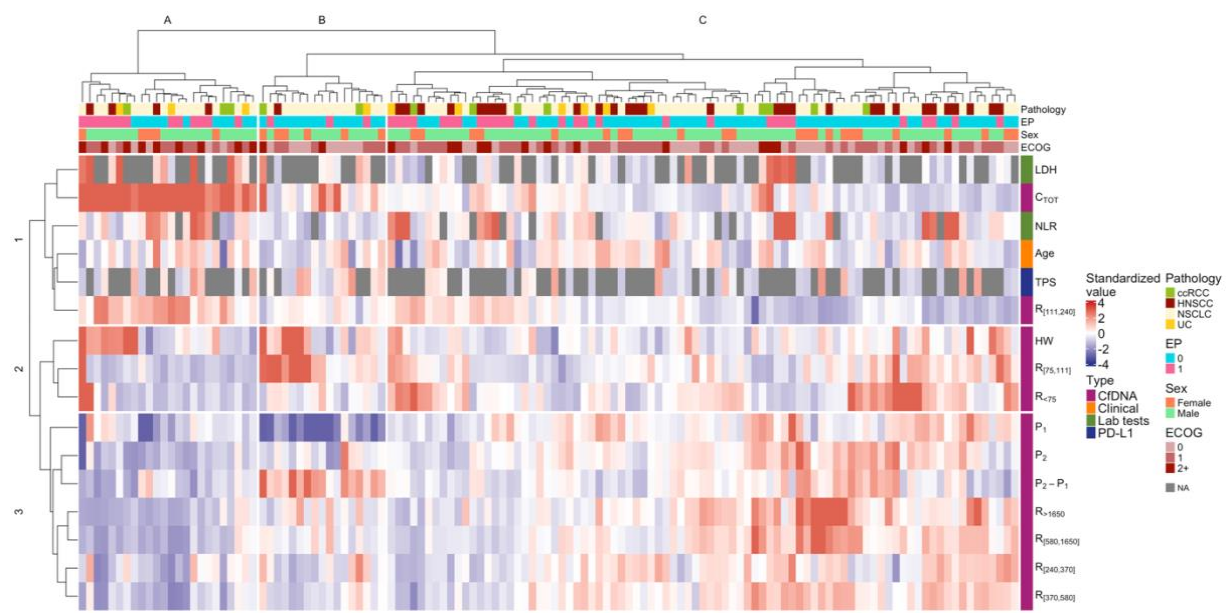

Two-way hierarchical clustering was computed to cluster patients (columns) and features (rows) simultaneously. The resulting dendrogram and clustered data are presented as a heatmap, where color intensity represents standardized feature values and the hierarchical relationships are shown by the accompanying dendrograms.

The three main feature groups were labeled from 1 to 3, while the three main patient groups were labeled from A to C.

(ccRCC: clear cell renal cell carcinoma; ECOG: Eastern Cooperative Oncology Group status; EP: early progression; HNSCC: head and neck squamous cell carcinoma; LDH: lactate dehydrogenase; NLR: neutrophil to lymphocyte ratio; NSCLC: non-small cell lung cancer; PD-L1: programmed cell death protein 1; TPS: tumor proportion score; UC: urothelial carcinoma)

*Figure 4: Higher ratio of long cfDNA fragments is associated with extended progression-free survival and no early progression under treatment in homogeneous subpopulations*

(A-D) CfDNA variables distribution are displayed using boxplots stratified by early progression status and Kaplan-Meier progression-free survival curves. Only the best cfDNA variables resulting from statistical analyses were displayed: first-line non-small cell lung cancer (NSCLC) patients: $P_2$, $P_2 - P_1$, $R_{>1650}$, and $C_{TOT}$; head and neck squamous cell carcinoma (HNSCC) patients treated with nivolumab: $R_{>1660}$ and $C_{TOT}$. The boxplots display the Welch two sample t-test significance. Kaplan-Meier plots display the log-rank test p-value, and the optimal threshold computed to stratify patients between long and short progression-free survival (see Methods). (B-E) Pearson correlation coefficients were computed between the eleven cfDNA variables and other clinical or biological features. Five of them were analyzed across both subpopulations: age, Eastern Cooperative Oncology Group (ECOG) status, neutrophil-to-lymphocyte ratio (NLR), lactate dehydrogenase (LDH), sum of the largest target diameters (SLD) (first-line NSCLC patients: $n_{SLD} = 58$, $n_{NLR} = 52$, $n_{LDH} = 16$; HNSCC nivolumab patients: $n_{SLD} = 22$, $n_{NLR} = 24$, $n_{LDH} = 24$). Additionally, tumor proportion score (TPS) was included for first-line NSCLC patients ($n_{TPS} = 55$) and combined positive score (CPS) was included for HNSCC nivolumab patients ($n_{CPS} = 10$). (C-F) Venn diagrams were generated based on the counts of true early progressors (pink) and true responders (blue) identified by the two best-performing cfDNA variables from the statistical analyses of each cohort and NLR.

(Significance: ****: p-value < 0.0001; ***: p-value < 0.001; **: p-value < 0.01; *: p-value < 0.05; n.s.: non-significant, > 0.05)

# TABLES

*Table 1: Clinical characteristics of the SChISM patients*

| Variable | Overall N = 126 [1] | Pathology UC N = 10 [1] | Pathology HNSCC N = 34 [1] | Pathology ccRCC N = 14 [1] | Pathology NSCLC N = 68 [1] | p-value[2] |
|---|---|---|---|---|---|---|
| **Sex** | | | | | | 0,6 |
| Male | 92 (73%) | 9 (90%) | 26 (76%) | 10 (71%) | 47 (69%) | |
| Female | 34 (27%) | 1 (10%) | 8 (24%) | 4 (29%) | 21 (31%) | |
| **Age (years)** | 64 (11) | 69 (8) | 64 (10) | 59 (14) | 63 (10) | 0,2 |
| **Treatment[3]** | | | | | | |
| Atezolizumab | 1 (0,8%) | 0 (0%) | 0 (0%) | 0 (0%) | 1 (1,5%) | |
| Nivolumab | 32 (25%) | 0 (0%) | 25 (74%) | 1 (7,1%) | 6 (8,8%) | |
| Nivolumab + Ipilimumab | 2 (1,6%) | 0 (0%) | 0 (0%) | 2 (14%) | 0 (0%) | |
| Nivolumab + targeted therapy | 4 (3,2%) | 0 (0%) | 0 (0%) | 4 (29%) | 0 (0%) | |
| Pembrolizumab | 32 (25%) | 9 (90%) | 3 (8,8%) | 2 (14%) | 18 (26%) | |
| Pembrolizumab + chemotherapy | 48 (38%) | 0 (0%) | 5 (15%) | 0 (0%) | 43 (63%) | |
| Pembrolizumab + targeted therapy | 7 (5,6%) | 1 (10%) | 1 (2,9%) | 5 (36%) | 0 (0%) | |
| **Treatment line** | | | | | | <0,001 |
| 1 | 100 (79%) | 3 (30%) | 23 (68%) | 14 (100%) | 60 (88%) | |
| 2 | 25 (20%) | 7 (70%) | 11 (32%) | 0 (0%) | 7 (10%) | |
| 3 | 1 (0,8%) | 0 (0%) | 0 (0%) | 0 (0%) | 1 (1,5%) | |
| **CPS (%)** | | | | | | 0,6 |
| [1, 20[ | 8 (40%) | 0 (0%) | 8 (42%) | 0 (-%) | 0 (-%) | |
| < 1 | 5 (25%) | 0 (0%) | 5 (26%) | 0 (-%) | 0 (-%) | |
| >= 20 | 7 (35%) | 1 (100%) | 6 (32%) | 0 (-%) | 0 (-%) | |
| Missing | 106 | 9 | 15 | 14 | 68 | |
| **TPS (%)** | | | | | | >0,9 |
| [1, 50[ | 25 (40%) | 0 (-%) | 0 (-%) | 0 (-%) | 25 (40%) | |
| < 1 | 14 (23%) | 0 (-%) | 0 (-%) | 0 (-%) | 14 (23%) | |
| >= 50 | 23 (37%) | 0 (-%) | 0 (-%) | 0 (-%) | 23 (37%) | |
| Missing | 64 | 10 | 34 | 14 | 6 | |
| **NLR** | 4,5 (3,8) | 3,6 (2,3) | 7,3 (5,4) | 2,5 (1,3) | 3,6 (2,3) | <0,001 |
| Missing | 12 | 0 | 2 | 1 | 9 | |
| **LDH (UI/L)** | 238 (147) | 221 (91) | 236 (195) | 252 (92) | 243 (88) | 0,4 |
| Missing | 56 | 0 | 1 | 3 | 52 | |
| **SLD (mm)** | 63 (50) | 50 (46) | 40 (22) | 92 (62) | 70 (53) | 0,016 |
| Missing | 7 | 0 | 5 | 0 | 2 | |
| **Mutations[4]** | | | | | | |
| KRAS | 26 (62%) | 0 (0%) | 0 (0%) | 0 (-%) | 26 (65%) | |
| TP53 | 25 (66%) | 1 (100%) | 1 (100%) | 0 (-%) | 23 (64%) | |
| **Early progression[5]** | | | | | | <0,001 |
| 0 | 82 (65%) | 3 (30%) | 14 (41%) | 11 (79%) | 54 (79%) | |
| 1 | 44 (35%) | 7 (70%) | 20 (59%) | 3 (21%) | 14 (21%) | |
| **PFS (median)** | 9,5 (6,8, —) | 2,8 (2,4, 5,6) | 2,6 (1,8, —) | 9,6 (6,0, —) | — (13, —) | <0,001 |

[1] n (%); Mean (SD)
[2] Fisher's exact test; Kruskal-Wallis rank sum test; NA
[3] Immunotherapy combination and potential associated therapy
[4] Other hidden mutations are: EGFR (3), BRAF (5), CTNNB1 (1), ERBB4 (1), FGFR 3 (1), P13 KINASE (1), PDGFRA (1), PIK3CA (2), PTEN (4), RET (1), STK11 (1), TTF1 (1)
[5] 1, progression before 3 months of immunotherapy | 0, otherwise

(ccRCC: clear cell renal cell carcinoma; CPS: combined positive score; HNSCC: head and neck squamous cell carcinoma; KRAS: Kirsten rat sarcoma viral oncogene; LDH: lactate dehydrogenase; NLR: neutrophil to lymphocyte ratio; NSCLC: non-small cell lung cancer; SLD: sum of the largest diameters of target lesions; TPS: tumor proportion score; TP53: transformation-related protein 53; UC: urothelial carcinoma)

*Table 2: Logistic and Cox regression results in the three cohorts*

(A)

**ALL PATIENTS: CLINICAL AND BIOLOGICAL VARIABLES**

| VARIABLE | LEVEL | N | EARLY PROGRESSION | | | | | PROGRESSION-FREE SURVIVAL | | | | |
|---|---|---|---|---|---|---|---|---|---|---|---|---|
| | | | AUC | OR UV | SIGNIF UV | OR MV | SIGNIF MV | C INDEX | HR UV | SIGNIF UV | HR MV | SIGNIF MV |
| CFDNA METRICS | | | | | | | | | | | | |
| $R_{>1650}$ | | 126 | 0.73 | 0.39 (0.25 – 0.62) | **** | 0.28 (0.13 – 0.6) | *** | 0.69 | 0.54 (0.42 – 0.68) | **** | 0.44 (0.33 – 0.58) | **** |
| $P_2 - P_1$ | | 126 | 0.72 | 0.38 (0.23 – 0.62) | *** | 0.36 (0.18 – 0.73) | ** | 0.67 | 0.55 (0.42 – 0.72) | **** | 0.6 (0.44 – 0.83) | ** |
| $R_{[580,1650]}$ | | 126 | 0.65 | 0.51 (0.32 – 0.82) | ** | 0.48 (0.23 – 0.99) | * | 0.64 | 0.55 (0.4 – 0.76) | *** | 0.52 (0.36 – 0.76) | *** |
| $P_2$ | | 126 | 0.64 | 0.56 (0.36 – 0.86) | ** | 0.46 (0.25 – 0.85) | * | 0.63 | 0.6 (0.46 – 0.79) | *** | 0.59 (0.44 – 0.8) | *** |
| $C_{TOT}$ | | 126 | 0.65 | 1.7 (1.1 – 2.5) | ** | 2 (1.1 – 3.4) | * | 0.62 | 1.5 (1.2 – 1.9) | *** | 1.8 (1.3 – 2.3) | **** |
| $R_{[111,240]}$ | | 126 | 0.60 | 1.5 (1 – 2.2) | * | 1.7 (0.94 – 3) | n.s. | 0.59 | 1.5 (1.2 – 2) | ** | 1.7 (1.3 – 2.4) | *** |
| $R_{[75,111]}$ | | 126 | 0.61 | 0.68 (0.46 – 0.99) | * | 0.76 (0.44 – 1.3) | n.s. | 0.59 | 0.75 (0.58 – 0.97) | * | 0.78 (0.59 – 1) | n.s. |
| $R_{[370,580]}$ | | 126 | 0.57 | 0.74 (0.51 – 1.1) | n.s. | 0.5 (0.27 – 0.94) | * | 0.57 | 0.73 (0.56 – 0.94) | * | 0.6 (0.44 – 0.82) | ** |
| $R_{[240,370]}$ | | 126 | 0.51 | 1 (0.69 – 1.4) | n.s. | 0.89 (0.52 – 1.5) | n.s. | 0.51 | 0.92 (0.71 – 1.2) | n.s. | 0.73 (0.53 – 0.99) | * |
| CLINICAL AND BIOLOGICAL VARIABLES | | | | | | | | | | | | |
| Neutrophil-to-lymphocyte ratio | | 114 | 0.69 | 2.4 (1.5 – 4) | *** | | | 0.63 | 1.6 (1.2 – 2.2) | ** | | |
| ECOG | 0 (reference) | 126 | | | | | | | | | | |
| | 2_sup | | | 6 (2.1 – 17) | *** | | | 0.62 | 3.4 (1.8 – 6.3) | **** | | |
| Pathology | NSCLC (reference) | 126 | | | | | | | | | | |
| | HNSCC | | | 5.5 (2.2 – 14) | *** | | | 0.60 | 3.8 (2.2 – 6.6) | **** | | |
| | UC | | | 9 (2.1 – 39) | ** | | | 0.60 | 2.9 (1.2 – 6.7) | * | | |
| Sex | male (reference) | 126 | | | | | | | | | | |
| | female | | | 0.12 (0.034 – 0.42) | *** | | | 0.59 | 0.28 (0.14 – 0.59) | *** | | |

(B)

**FIRST LINE NSCLC: CLINICAL AND BIOLOGICAL VARIABLES**

| VARIABLE | LEVEL | N | EARLY PROGRESSION | | | | | PROGRESSION-FREE SURVIVAL | | | | |
|---|---|---|---|---|---|---|---|---|---|---|---|---|
| | | | AUC | OR UV | SIGNIF UV | OR MV | SIGNIF MV | C INDEX | HR UV | SIGNIF UV | HR MV | SIGNIF MV |
| CFDNA METRICS | | | | | | | | | | | | |
| $P_2$ | | 60 | 0.918 | 0.067 (0.011 – 0.41) | ** | 0.0001 (1.2e-07 – 9.9) | n.s. | 0.76 | 0.33 (0.2 – 0.54) | **** | 0.4 (0.23 – 0.7) | ** |
| $P_2 - P_1$ | | 60 | 0.820 | 0.19 (0.059 – 0.61) | ** | 0.25 (0.039 – 1.6) | n.s. | 0.75 | 0.39 (0.23 – 0.66) | *** | 0.47 (0.26 – 0.85) | * |
| $R_{>1650}$ | | 60 | 0.840 | 0.22 (0.091 – 0.56) | ** | 0.0012 (1.4e-10 – 11000) | n.s. | 0.74 | 0.41 (0.27 – 0.64) | **** | 0.37 (0.21 – 0.67) | *** |
| $C_{TOT}$ | | 60 | 0.830 | 4 (1.6 – 9.9) | ** | 4.8 (0.66 – 35) | n.s. | 0.71 | 2.4 (1.6 – 3.6) | **** | 1.9 (1.2 – 3.1) | ** |
| $R_{[370,580]}$ | | 60 | 0.830 | 0.19 (0.061 – 0.62) | ** | 0.26 (0.044 – 1.5) | n.s. | 0.71 | 0.42 (0.24 – 0.74) | ** | 0.52 (0.28 – 0.93) | * |
| $R_{[580,1650]}$ | | 60 | 0.780 | 0.21 (0.059 – 0.73) | * | 0.55 (0.097 – 3.1) | n.s. | 0.70 | 0.38 (0.19 – 0.75) | ** | 0.52 (0.26 – 1) | n.s. |
| $R_{[111,240]}$ | | 60 | 0.750 | 2.9 (1.2 – 6.9) | * | 1.6 (0.37 – 6.9) | n.s. | 0.67 | 2.2 (1.3 – 3.6) | ** | 1.7 (0.97 – 3) | n.s. |
| CLINICAL AND BIOLOGICAL VARIABLES | | | | | | | | | | | | |
| Neutrophil-to-lymphocyte ratio | | 54 | 0.790 | 6.1 (1.4 – 26) | * | | | 0.67 | 1.6 (0.85 – 3.1) | n.s. | | |
| ECOG | 0 (reference) | 60 | | | | | | | | | | |
| | 2_sup | | | 14 (1.6 – 120) | * | | | 0.62 | 3.8 (1.2 – 13) | * | | |
| Sex | male (reference) | 60 | | | | | | | | | | |
| | female | | | 0.27 (0.031 – 2.4) | n.s. | | | 0.62 | 0.12 (0.017 – 0.94) | * | | |

(C)

**HNSCC TREATED WITH NIVOLUMAB: CLINICAL AND BIOLOGICAL VARIABLES**

| VARIABLE | LEVEL | N | EARLY PROGRESSION | | | | | PROGRESSION-FREE SURVIVAL | | | | |
|---|---|---|---|---|---|---|---|---|---|---|---|---|
| | | | AUC | OR UV | SIGNIF UV | OR MV | SIGNIF MV | C INDEX | HR UV | SIGNIF UV | HR MV | SIGNIF MV |
| CFDNA METRICS | | | | | | | | | | | | |
| $R_{>1650}$ | | 25 | 0.75 | 0.39 (0.12 – 1.3) | n.s. | 0.00035 (8.9e-15 – 1.4e+07) | n.s. | 0.67 | 0.58 (0.38 – 0.9) | * | 0.41 (0.23 – 0.75) | ** |
| $R_{[580,1650]}$ | | 25 | 0.62 | 0.56 (0.22 – 1.4) | n.s. | 0.28 (0.024 – 3.2) | n.s. | 0.60 | 0.63 (0.39 – 1) | n.s. | 0.56 (0.32 – 0.98) | * |
| CLINICAL AND BIOLOGICAL VARIABLES | | | | | | | | | | | | |
| Neutrophil-to-lymphocyte ratio | | 24 | 0.80 | 3.2 (1 – 9.8) | * | | | 0.58 | 1.5 (0.9 – 2.4) | n.s. | | |
| Sex | male (reference) | 25 | | | | | | | | | | |
| | female | | | 0.078 (0.0064 – 0.97) | * | | | 0.56 | 0.5 (0.15 – 1.7) | n.s. | | |

(A) Multi-cancer cohort results (B) First-line NSCLC cohort results (C) Nivolumab-treated HNSCC cohort results

(AUC: area under the receiving operator curve; ECOG: Eastern Cooperative Oncology Group status; HNSCC: head and neck squamous cell carcinoma; HR: hazard ratio; MV: multivariable; NSCLC: non-small cell lung cancer; OR: odds ratio; UV: univariable)

(Significance: ****: p-value < 0.0001; ***: p-value < 0.001; **: p-value < 0.01; *: p-value < 0.05; n.s.: non-significant, > 0.05)

The categorical features present results for each of their levels comparing to the reference level.

# SUPPLEMENTARY

*Figure S1: Time-to-progression distributions vary across pathological subpopulations*

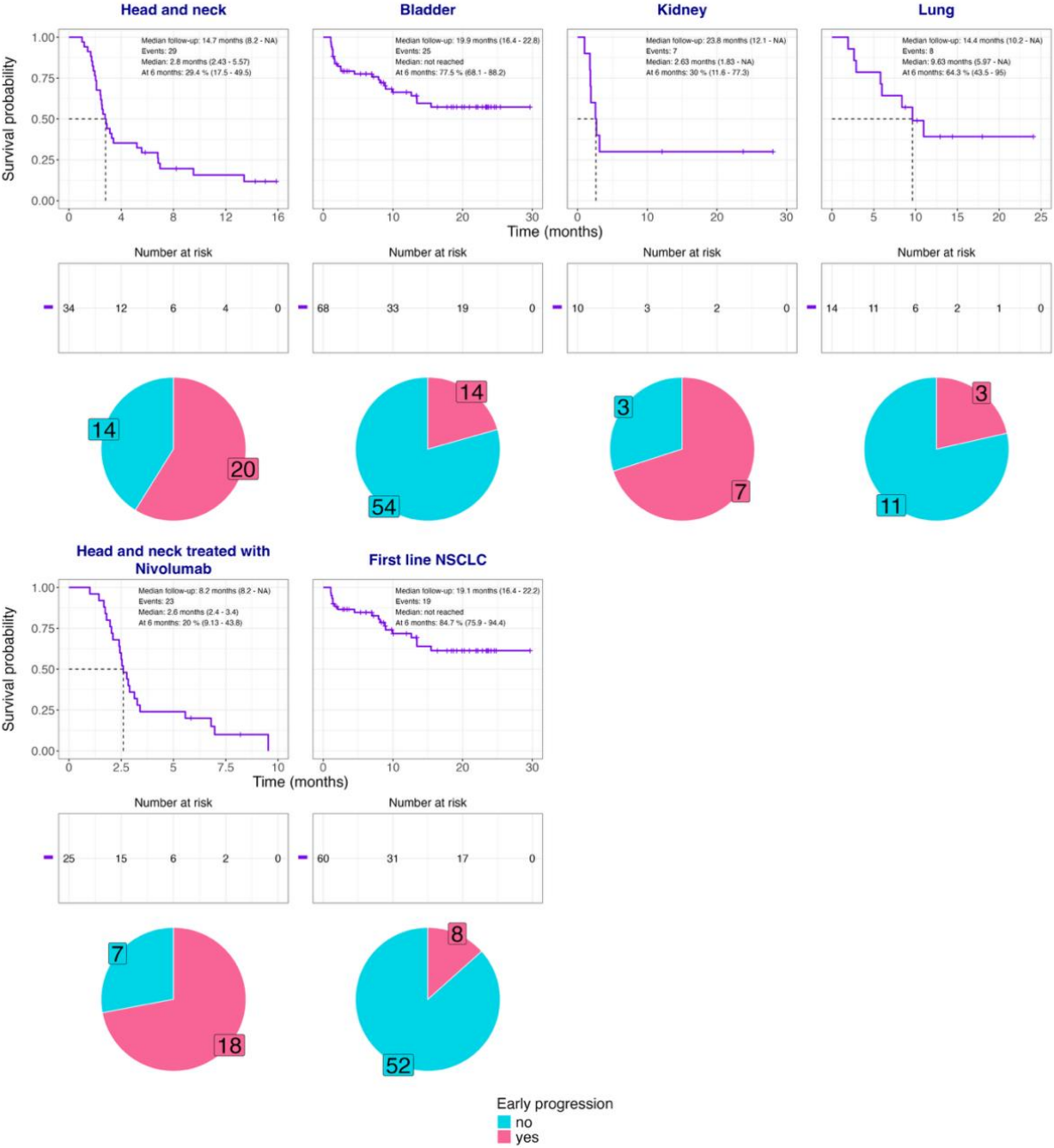

Survival curves (purple line) of the progression-free survival, estimated by Kaplan-Meier method, across all pathological subpopulations. Dotted lines indicate the median time of progression. Pie charts display the early-progression distribution across the same subpopulations.

*Figure S2: Time-to-death distributions across pathological subpopulations*

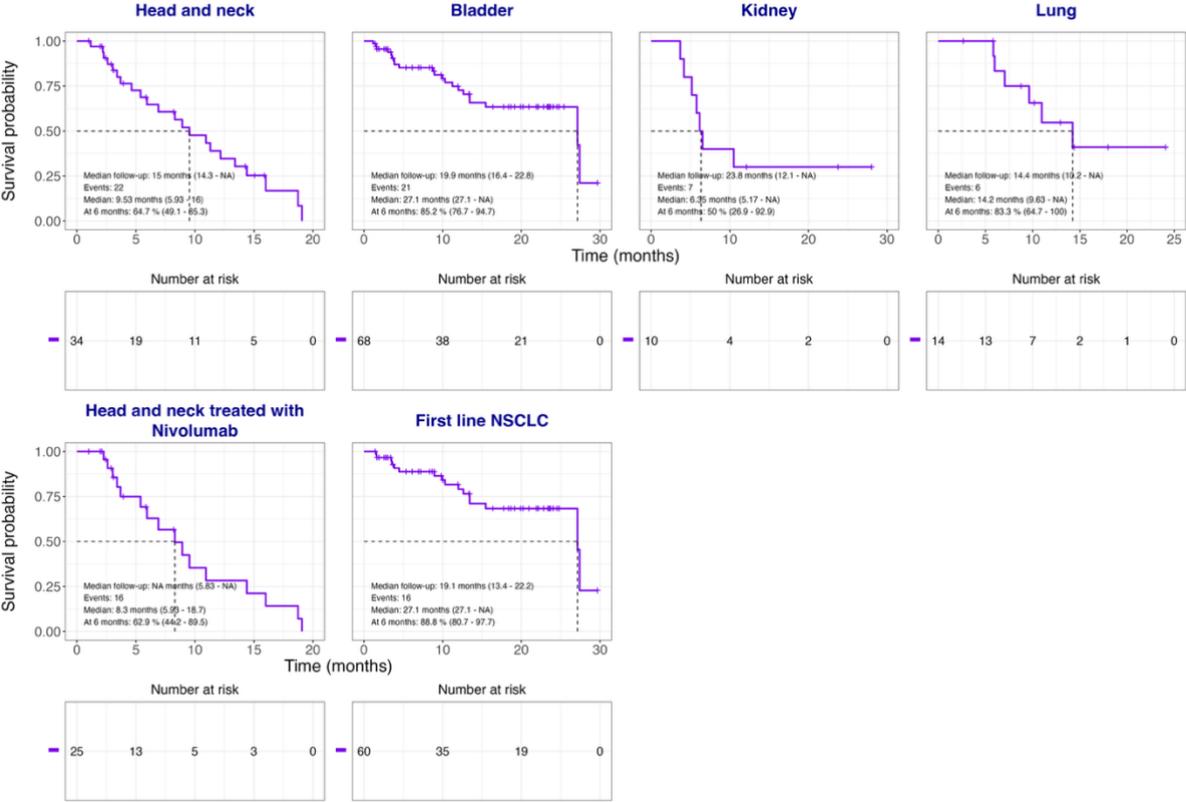

Survival curves (purple line) of the overall survival, estimated by Kaplan-Meier method, across different pathological subpopulations. Dotted lines indicate the median time of progression

*Figure S3: Distribution of cfDNA size profile variables*

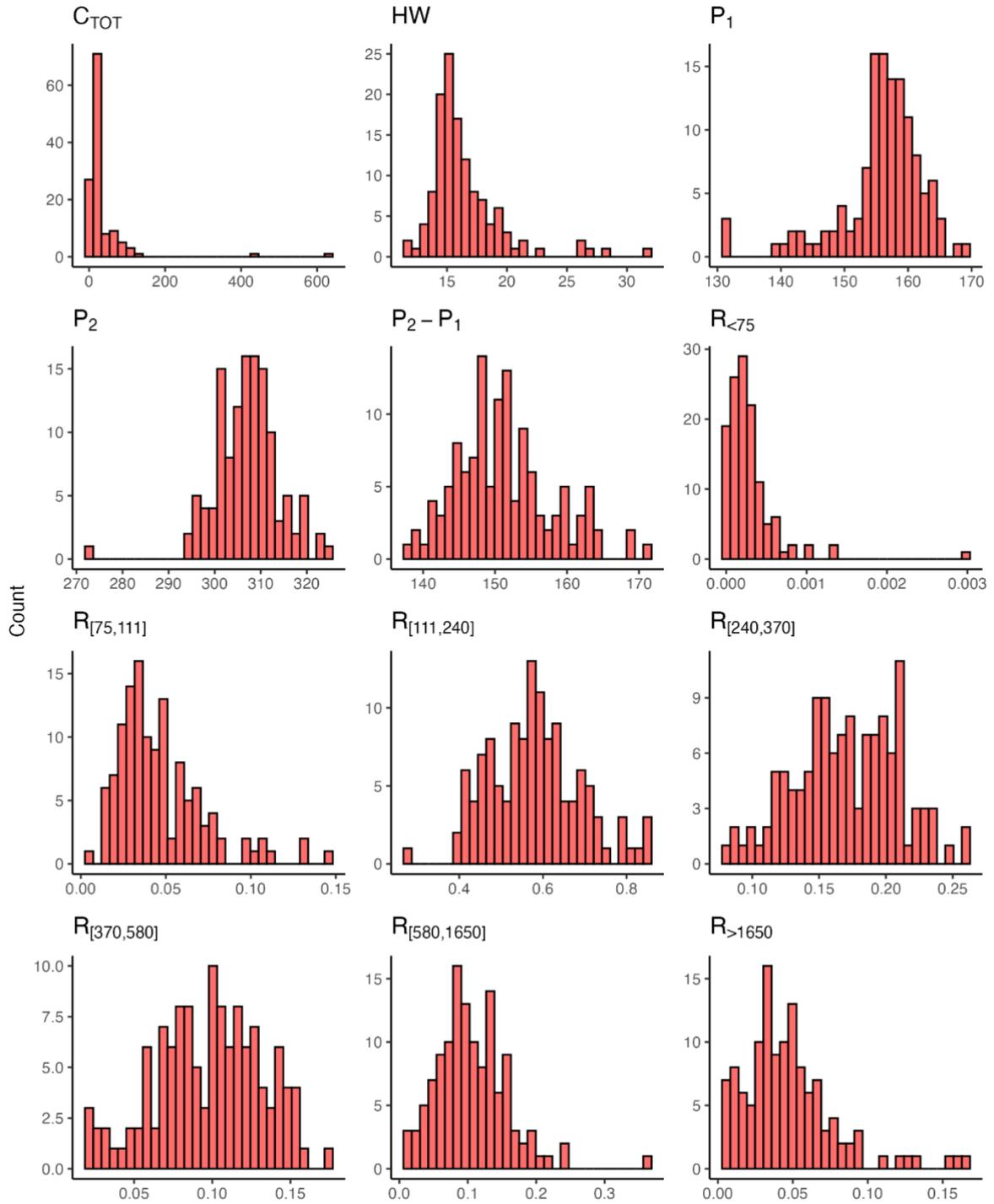

*Figure S4: Higher neutrophil-to-lymphocyte ratio is associated with shorter progression-free survival and early progression under treatment*

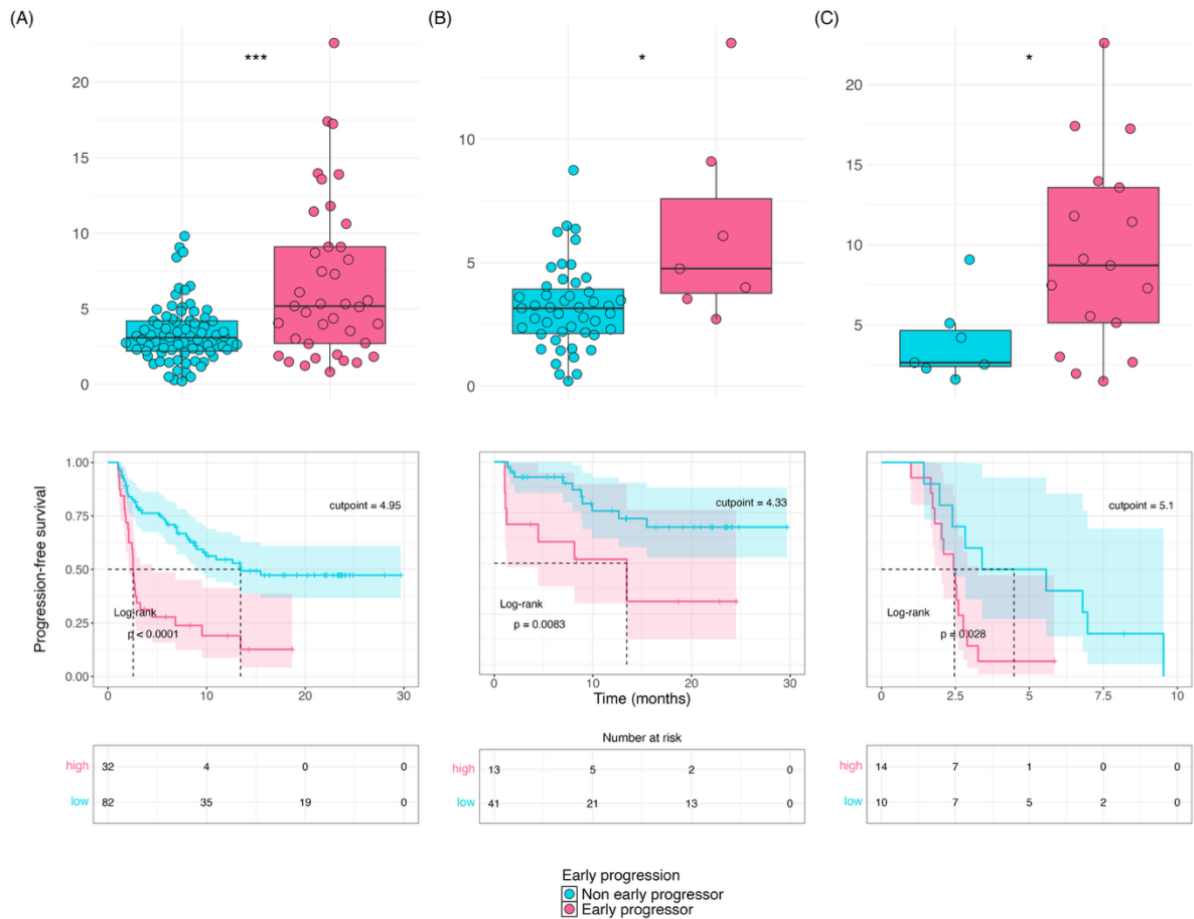

Neutrophil-to-lymphocyte ratio distribution is displayed using boxplots stratified by early progression status and Kaplan-Meier progression-free survival curves in the (A) multi-cancer cohort (A), firs-line non squamous cell lung cancer patients (B), and head and neck squamous cell carcinoma patients treated with nivolumab (C).

Boxplots display the Welch two sample t-test significance. Kaplan-Meier plots display the log-rank test p-value and the optimal threshold computed to stratify patients between long and short progression-free survival (see Methods).

(Significance: \*\*\*\*: p-value < 0.0001; \*\*\*: p-value < 0.001; \*\*: p-value < 0.01; \*: p-value < 0.05; n.s.: non-significant, > 0.05)

*Table s1: Empirical statistics of cfDNA variables*

| Variable | Overall, N = 126[1] | Early-progression Early-progressors N = 44[1] | Non early-progressors N = 82[1] | p-value[2] |
|---|---|---|---|---|
| $C_{TOT}$ | 16 [6-637] | 18 [7-637] | 14 [6-109] | 0,006 |
| $HW$ | 15.61 [11.84-31.82] | 15.98 [12.82-28.12] | 15.50 [11.84-31.82] | 0,4 |
| $P_1$ | 157 [132-169] | 157 [132-169] | 156 [132-168] | 0,3 |
| $P_2$ | 307 [274-326] | 306 [274-319] | 308 [295-326] | 0,010 |
| $P_2 - P_1$ | 151 [138-172] | 148 [138-159] | 152 [139-172] | <0,001 |
| $R_{<75}$ | 0.0002 [0.0000-0.0030] | 0.0002 [0.0000-0.0014] | 0.0002 [0.0000-0.0030] | 0,8 |
| $R_{[75.111]}$ | 0.040 [0.006-0.147] | 0.034 [0.006-0.129] | 0.044 [0.013-0.147] | 0,041 |
| $R_{[111.240]}$ | 0.58 [0.28-0.85] | 0.60 [0.41-0.85] | 0.57 [0.28-0.82] | 0,055 |
| $R_{[240.370]}$ | 0.17 [0.08-0.26] | 0.17 [0.08-0.26] | 0.17 [0.09-0.26] | 0,8 |
| $R_{[370.580]}$ | 0.10 [0.02-0.17] | 0.09 [0.02-0.15] | 0.10 [0.03-0.17] | 0,2 |
| $R_{[580.1650]}$ | 0.10 [0.01-0.36] | 0.08 [0.01-0.17] | 0.11 [0.02-0.36] | 0,005 |
| $R_{>1650}$ | 0.043 [0.005-0.165] | 0.031 [0.005-0.086] | 0.050 [0.006-0.165] | <0,001 |

[1] Median [Min-Max]
[2] Wilcoxon rank sum test

## ACKNOWLEDGEMENTS

This work received support from the French government under the France 2030 investment plan, as part of the Initiative d'Excellence d'Aix-Marseille Université - A*MIDEX (AMX-19-IET-001 & AMX-21-IET-017).

## Declaration of Interest statement

FF, FG and AB are all members of Adelis Technologies, Grabels, France. FF is member of IDS-Solutions Oncology (Marseille, France).

8. Mattox AK, Douville C, Wang Y, et al. The Origin of Highly Elevated Cell-Free DNA in Healthy Individuals and Patients with Pancreatic, Colorectal, Lung, or Ovarian Cancer. *Cancer Discov*. 2023;13(10):2166-2179. doi:10.1158/2159-8290.CD-21-1252

9. Heitzer E, Auinger L, Speicher MR. Cell-Free DNA and Apoptosis: How Dead Cells Inform About the Living. *Trends in Molecular Medicine*. 2020;26(5):519-528. doi:10.1016/j.molmed.2020.01.012

10. Hu Z, Chen H, Long Y, Li P, Gu Y. The main sources of circulating cell-free DNA: Apoptosis, necrosis and active secretion. *Crit Rev Oncol Hematol*. 2021;157:103166. doi:10.1016/j.critrevonc.2020.103166

11. Thakur BK, Zhang H, Becker A, et al. Double-stranded DNA in exosomes: a novel biomarker in cancer detection. *Cell Res*. 2014;24(6):766-769. doi:10.1038/cr.2014.44

12. Khier S, Lohan L. Kinetics of circulating cell-free DNA for biomedical applications: critical appraisal of the literature. *Future Science OA*. 2018;4(4):FSO295. doi:10.4155/fsoa-2017-0140

13. Esposito A, Bardelli A, Criscitiello C, et al. Monitoring tumor-derived cell-free DNA in patients with solid tumors: clinical perspectives and research opportunities. *Cancer Treat Rev*. 2014;40(5):648-655. doi:10.1016/j.ctrv.2013.10.003

14. Thierry AR, El Messaoudi S, Gahan PB, Anker P, Stroun M. Origins, structures, and functions of circulating DNA in oncology. *Cancer Metastasis Rev*. 2016;35(3):347-376. doi:10.1007/s10555-016-9629-x

15. Stadler JC, Belloum Y, Deitert B, et al. Current and Future Clinical Applications of ctDNA in Immuno-Oncology. *Cancer Research*. 2022;82(3):349-358. doi:10.1158/0008-5472.CAN-21-1718

16. Ivanov M, Baranova A, Butler T, Spellman P, Mileyko V. Non-random fragmentation patterns in circulating cell-free DNA reflect epigenetic regulation. *BMC Genom*. 2015;16(Suppl 13):S1. doi:10.1186/1471-2164-16-S13-S1

17. Mouliere F, Chandrananda D, Piskorz AM, et al. Enhanced detection of circulating tumor DNA by fragment size analysis. *Sci Transl Med*. 2018;10(466):eaat4921. doi:10.1126/scitranslmed.aat4921

18. Qi T, Pan M, Shi H, Wang L, Bai Y, Ge Q. Cell-Free DNA Fragmentomics: The Novel Promising Biomarker. *Int J Mol Sci*. 2023;24(2):1503. doi:10.3390/ijms24021503

39. Sicard G, Fina F, Fanciullino R, Barlesi F, Ciccolini J. Like a Rolling Stone: Sting-Cgas Pathway and Cell-Free DNA as Biomarkers for Combinatorial Immunotherapy. *Pharmaceutics*. 2020;12(8):758. doi:10.3390/pharmaceutics12080758

40. Grabuschnig S, Bronkhorst AJ, Holdenrieder S, et al. Putative Origins of Cell-Free DNA in Humans: A Review of Active and Passive Nucleic Acid Release Mechanisms. *Int J Mol Sci*. 2020;21(21):8062. doi:10.3390/ijms21218062

41. Lo YMD, Han DSC, Jiang P, Chiu RWK. Epigenetics, fragmentomics, and topology of cell-free DNA in liquid biopsies. *Science*. 2021;372(6538):eaaw3616. doi:10.1126/science.aaw3616

42. Kwon J, Bakhoum SF. The cytosolic DNA-sensing cGAS-STING pathway in cancer. *Cancer Discov*. 2020;10(1):26-39. doi:10.1158/2159-8290.CD-19-0761

43. Luecke S, Holleufer A, Christensen MH, et al. cGAS is activated by DNA in a length-dependent manner. *EMBO reports*. 2017;18(10):1707-1715. doi:10.15252/embr.201744017

44. Fridlich O, Peretz A, Fox-Fisher I, et al. Elevated cfDNA after exercise is derived primarily from mature polymorphonuclear neutrophils, with a minor contribution of cardiomyocytes. *Cell Rep Med*. 2023;4(6):101074. doi:10.1016/j.xcrm.2023.101074

45. Mouliere F, Robert B, Peyrotte EA, et al. High Fragmentation Characterizes Tumour-Derived Circulating DNA. *PLOS ONE*. 2011;6(9):e23418. doi:10.1371/journal.pone.0023418

46. Benzekry S. Artificial Intelligence and Mechanistic Modeling for Clinical Decision Making in Oncology. *Clinical Pharmacology & Therapeutics*. 2020;108(3):471-486. doi:10.1002/cpt.1951